\documentclass[]{aa}
\usepackage{graphicx}
\usepackage{natbib}

\bibdata{lf1252}

\begin{document}

\offprints{S. Toft}

\title{$K_s$-band luminosity function of the $z=1.237$ cluster of galaxies \object{RDCS J1252.9$-$2927}\thanks{Based on observations obtained at the European Southern Observatory using the ESO Very Large Telescope on Cerro Paranal (ESO program 166.A-0101).} }
\author{S. Toft \inst{1,2}, V. Mainieri \inst{3}, P. Rosati \inst{4}, C. Lidman \inst{5}, R. Demarco \inst{6}, M. Nonino \inst{7}, S. A. Stanford \inst{8}$^,$\inst{9} }
\authorrunning{Toft et al.}

\institute{
Astronomical Observatory, University of Copenhagen, Juliane Maries Vej 30, DK-2100 Copenhagen \O, Denmark 
\and
Department of Astronomy, Yale University, P.O. Box 208101, New Haven, CT 06520-8101, USA\\ \email{toft@astro.yale.edu}
\and
Max-Planck-Institut f\"{u}r extraterrestrische Physik, Postfach 1319, D-85748 Garching, Germany 
\and
European Southern Observatory, Karl-Scwarzchild-Strasse 2, D-85748 Garching, Germany 
\and
European Southern Observatory, Alonso de Cordova 3107, Casilla 19001, Santiago, Chile 
\and 
Institute d'Astrophysique de Paris, 98bis Boulevard Arago, F-75014 Paris, France  
\and
Instituto Nazionale di Astrofisica, Osservatorio Astronomico de Trieste, via G.B. Tiepolo 11, I-34131, Trieste, Italy
\and
Department of Physics, University of California at Davis, 1 Shields Avenue, Davis, CA 95616, USA
\and
Institute of Geophysics and Planetary Physics, LLNL, Livermore, CA 94551,  USA
}

\abstract{
Using deep VLT/ISAAC near-infrared imaging data, we derive the $K_s$-band luminosity function (LF) of the $z=1.237$ massive X-ray luminous cluster of galaxies \object{RDCS J1252.9$-$2927}. 
Photometric redshifts, derived from deep multi-wavelength $BVRIzJK_s$ data, and calibrated using a large subset of galaxies with spectroscopic redshifts, are used to separate the cluster galaxy population from the foreground and background field galaxy population. This allows for a simultaneous determination of the characteristic magnitude $K^*_s$ and faint end slope $\alpha$ of the LF without having to make an uncertain statistical background subtraction.
The derived LF is well represented by the Schechter function with $K_s^*=18.54^{+0.45}_{-0.55}$ and $\alpha=-0.64^{+0.27}_{-0.25}$.
The shape of the bright end of the derived LF is similar to that measured at similar restframe wavelengths (in the $z$-band) in local clusters, but the characteristic magnitude is brighter by $\Delta M_z^*=-1.40^{+0.49}_{-0.58}$ magnitudes.
The derived $\alpha$ is similar to the value derived in the $K_s$-band in the $z=1$ cluster of galaxies \object{MG2016+112} but is shallower (at the $2.2\sigma$ level) than the value measured at similar restframe wavelength (in the $z$-band) in clusters in the local universe.
The brightening of the characteristic magnitude and lack of evolution in the shape of the bright end of the LF suggests that the massive cluster ellipticals 
that dominate the bright end of the LF were already in place at $z=1.237$, while the flattening of the faint end slope suggest that clusters at $z\sim 1$ contains relatively smaller fractions of low mass galaxies than clusters in the local universe.
The results presented in this paper are a challenge for semi analytical hierarchical models of galaxy formation which predict the characteristic magnitude to grow fainter and the faint end slope to steepen with redshift as the massive galaxies break up into their progenitors. 
The observed evolution is consistent with a scenario in which clusters are composed of a population of massive galaxies which formed at high redshift ($z\gg1$) and subsequently evolved passively, and a population of lower mass galaxies which are gradually accreted from the field, primarily at lower redshift ($z<1$).

}

\maketitle


\section{Introduction}

The galaxy luminosity function (LF) is 
a powerful tool for constraining models of galaxy formation and evolution. 
The LF of both field and cluster galaxies have been shown to be well represented by the Schechter function
$\phi(m){\rm{d}}m= n^*\left [ 10^{0.4(m^*-m)}\right ]^{\alpha+1}e^{-10^{0.4(m^*-m)}}{\rm{d}}m,$
where $m^*$ is the characteristic magnitude of the distribution, $\alpha$ is the faint end slope, and $n^*$ is a normalization constant describing the space density of galaxies.

The variation of the LF parameters with galaxy type (early-type/late-type), wavelength (UV, optical, near-infrared [NIR]) and environment (cluster versus field) and their evolution with redshift depends on the details of mass assembly and star formation and therefore provide strong constraints on galaxy evolution models.

At optical wavelengths the LF of field galaxies and cluster galaxies have been derived from a number of studies \citep[e.g.][]{garilli1999,blanton2001,goto02,blanton2003,christlein2003,depropris03}. The resulting LF parameters varies somewhat from study to study, most likely due to differences in the data quality, the employed methods for selecting complete samples galaxies and deriving their luminosity functions. 
 
The most extensive studies are derived from the Sloan Digital Sky Survey \citep{york2000}. Initial results based on the SDSS commissioning data \citep{blanton2001,goto02} disagreed with results from the literature, probably due to problems with the photometry, k-corrections and evolutionary corrections. Studies based on more recent data, however,  agree better with the literature and do not seem to have such problems \citep{blanton2003, popesso2004}. These studies show that the LF of field galaxies and cluster galaxies are very similar.
The LF in both cases has brighter characteristic magnitudes and steeper slopes in the redder bands than in the bluer bands. The faint end slope is slightly steeper for cluster galaxies than for field galaxies in all bands, while the characteristic magnitude from these studies appear to depend very little on environment. 
Rich evolved clusters with  cD galaxies  tend to have flatter faint end slopes than poorer clusters, indicating a deficiency of star forming dwarf galaxies in rich clusters \citep{lopez-cruz97,driver03}. \cite{Barkhouse} find a tendency for the faint end slope, of the LF of local clusters, to steepen with cluster centric distance,  consistent with the hypothesis that a large fraction of dwarf galaxies near cluster centers are being tidally disrupted.

While the variation in the LF parameters at optical wavelengths are sensitive to the star formation properties and morphological mix of the underlying galaxy population, the NIR LF is an excellent probe of the mass function of clusters. The $K$-band flux of a galaxy is a good tracer of its stellar mass \citep[e.g.][]{gavazzi96} it is not very sensitive to star formation and attenuation by dust (compared to the optical band) and the k-corrections are fairly small and relatively independent of galaxy type even at high redshifts \citep[e.g.][]{mannucci01}.
The evolution of the mass function is directly predicted from galaxy formation and evolution models,
making the $K$-band LF of field and cluster galaxies very useful for fundamental testing of the models.

Locally, the field galaxy $K$-band LF has been derived using the 2 Micron All Sky Survey \citep[2MASS,][]{jarrett00}. Early-type and late-type field galaxies have similarly shaped LF but the early-types are brighter (more massive) and less numerous than the late-types \citep{kochanek01}. 
The evolution of the field galaxy $K$-band LF has been estimated in three broad redshift bins centered on $z=0.5$, 1.0 and 1.5 using the K20 survey \citep{pozzetti03}. No evolution is found in $\alpha$  to $z=1.0$, in the highest redshift bin the data are  not deep enough to constrain it. 
The characteristic $K$-band magnitude of the galaxies in the  $z=1$ bin (which includes galaxies in the redshift range $0.75<z<1.3$) is consistent with a mild luminosity evolution in the $K$-band to $z\sim1$ of $\Delta M_K^*=-0.54\pm0.12$. 
At $z=1$, the bright end of the LF is still dominated by ellipticals, and only  a small decrease in their space density is observed, indicating that the field ellipticals are largely in place at $z=1$.

In a deep study of the  LF of the Coma cluster, \cite{depropris98coma} find a  characteristic magnitude $M_K^*=-24.02$ and a faint end slope of $\alpha=-0.78$. If they restrict the fit to ``bright'' galaxies ($M<M^*+3$) they find $\alpha=-0.9$.
\cite{depropris99} studied 38 clusters in the redshift range $0.1<z<0.9$ to look for an evolution in the $K$-band cluster LF. Their data was not deep enough to constrain the faint end slope , so they fixed it on the Coma cluster value $\alpha=-0.9$, and concentrated on searching for an evolution in the  properties of the bright end.
Using the same method as \cite{depropris99} the sample has subsequently been extended to include a few clusters in the redshift range $z=1.0-1.2$ \citep{kodama03,nakata01}.
The conclusion from these studies is that the shape of the bright end of the $K$-band LF appears unchanged to $z=1.2$ and the evolution of the characteristic $K$-band magnitude with redshift $K^*(z)$ is well described by passively evolving  galaxies assembled  at $z_f>2$. 

A high formation redshift is also inferred from the evolution of the colour magnitude (CM) relation of cluster early-types to $z\sim1.3$ \citep{bower92,aragon-salamanca93,SED98,stanford97,rosati99,vandokkum01,NIRpaper}. The simplest explanation for the evolution of the CM relation is the monolithic collapse models \citep[e.g.][]{eggen} in which the bulk of stars are formed in a single short duration burst at high redshift and subsequently evolve passively, but it can be equally well explained by semi analytical hierarchical models \citep[e.g.][]{kauffmann98} in which the stars are formed in the disks of of late-type progenitors which later merge to form the elliptical galaxies, provided that the merging does not trigger significant star formation.  

The $K$-band LF is predicted to evolve quite differently in the two formation scenarios, due to its close relationship with the evolution of the mass function.

If the galaxies were assembled at high redshift and subsequently evolved passively, as predicted by the monolithic models, 
we expect the characteristic magnitude to brighten with redshift (as a consequence of the passive evolution of the stars) but the shape of the LF to remain the same. 
If the galaxies were build up through continuous merging over a broader redshift interval as predicted in  hierarchical models, 
we expect the shape of the LF to change with redshift to reflect the accretion history of the cluster and the merging history of the cluster galaxies. 
A deficiency of the brightest galaxies should appear as the the most massive galaxies  break up into their progenitors, resulting in a steepening of the bright end of the LF with redshift. 
If small structures form first and subsequently merge to form larger structures, the faint end slope should steepen with redshift as galaxies break up into their progenitors. In many hierarchical models however, the cluster population grow by accretion of surrounding structures. The accretion history, and the  mass spectrum of the accreted structures could influence the evolution of the LF.
The Butcher-Oemler effect \citep{butcher78} in intermediate redshift clusters is probably due to star formation  in low mass galaxies which have been accreted from the field, induced by interaction with the cluster environment \citep{butcher84,fabricant91,smail98,vandokkum00,depropris03}. Such an accretion of primarily low mass galaxies could cause the faint end slope to decline with redshift, as a smaller number of low mass galaxies would have had time to be accreted onto the cluster environment at high redshift.

The apparent passive evolution of the bright end of the cluster LF to $z\sim1$  is a challenge for hierarchical models which predict that the characteristic  mass  should be a factor of three smaller at $z \sim 1$ than at present \citep{kodama03}.    
None of the studies mentioned above have been able to constrain the evolution of $\alpha$.
The first constraints on $\alpha$ at high redshift was derived by \cite{toft2016paper} who  simultaneously fitted $K^*$ and $\alpha$ in the $z=1$ cluster MG2016+112. The derived constraints on $\alpha$ are not strong ($\alpha=-0.60^{+0.39}_{-0.33}$), however it is important to keep it free in the fit since it is coupled to the derived value of $K^*$ and its uncertainties. Furthermore it is noted that since $\alpha$ depends on wavelength, and the $K$-band corresponds roughly to restframe $z$-band at $z \sim 1$, the derived value should be compared to the local value measured in the $z$-band rather than in the $K$-band when studying evolutionary patterns.  

In this paper, we derive firm constraints on $K^*$ and $\alpha$ for the $K$-band LF of the massive, X-ray luminous cluster of galaxies \object{RDCS J1252.9-2927} \citep{xpaper} at $z=1.237$.  We apply the method of \cite{toft2016paper} to high quality multi-wavelength NIR and optical photometric data and extensive spectroscopic follow up observations covering a $4\arcmin \times 4 \arcmin$ field around the cluster.

The outline of this paper is as follows:
In \S\ref{sec.sample} we give a brief introduction to the data and describe how cluster galaxies are separated from the field galaxy population using photometric redshifts, calibrated using the  spectroscopic data. In \S\ref{sec.lumfun} we derive the LF of the cluster galaxies, in \S\ref{sec.evolution} we  compare to results from studies of lower redshift clusters to look for an evolution and compare it to model predictions, and in \S\ref{sec.discussion} we summarize and discuss the results.
Throughout the paper we assume a flat cosmology, with $\Omega_m=0.3$, $\Omega_{\Lambda}=0.7$ and $H_0=70$ km\,s$^{-1}$\,Mpc$^{-1}$. Magnitudes are in the the Vega system.

\section{Building a complete sample of cluster galaxies}
\label{sec.sample}
The observational basis of this study consist of high quality multi-wavelength NIR  and optical  photometric data of a $4\arcmin \times 4 \arcmin$ field around \object{RDCS J1252.9-2927}, supplemented by spectroscopic follow up observations, all obtained at the ESO Very Large Telescope (with the ISAAC, FORS1 and FORS2 instruments).  
The central part of the field was imaged for 24 hours in each of the NIR wavebands: $J_s$ and $K_s$, and for approximately 1 hour in each of the optical wavebands: $B$, $V$, $R$, $I$ and $z$.  
The full data set is described in detail in three companion papers: a paper describing the NIR data \citep{NIRpaper} a paper describing the optical data \citep{opticalpaper} and a paper describing the spectroscopic data \citep{specpaper}.

To derive the luminosity function of the cluster galaxies in \object{RDCS J1252.9-2927} we need a catalog of cluster galaxies which is complete to a certain magnitude. 
In Fig.~\ref{complete} we plot a binned representation of the number of detected objects in the $K_s$-band as function of magnitude \citep{NIRpaper}. From visual inspection of the turnover magnitude in this figure we estimate that the $K_s$-band data are complete to $K_s=22.5$.  
\begin{figure}
\setcounter{figure}{0}
\resizebox{\hsize}{!}{{\includegraphics{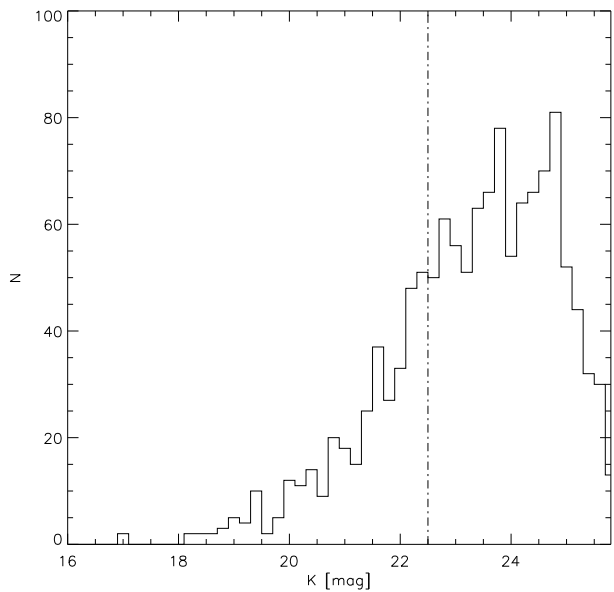}}}
\caption{Binned representation (bin size 0.2 mag) of the number of detected objects as a function of $K_s$-band magnitude. The data are estimated to be complete to $K_s=22.5$.}
\label{complete}
\end{figure}
To build a sample of cluster galaxies which is complete to this magnitude, the redshift distribution of all galaxies in the field must be determined in order to separate the cluster galaxy population from the foreground and background populations of galaxies. 

At faint magnitudes it is not feasible to obtain this solely through spectroscopic observations. 
Instead we take advantage of the extensive optical and NIR imaging data to derive photometric redshifts and  calibrate and test these against a  large subset of galaxies with spectroscopic redshifts.     

\subsection{Photometry and photometric redshifts}
We used SExtractor \citep{bertin} for object 
detection in each of the available wave bands, for computing magnitudes in apertures of $2{\arcsec}$ and for cross-correlation of the resulting catalogues. 
The ($3 \sigma$) limiting (Vega) magnitudes achieved for the imaging in $2^{\prime \prime}$ apertures are:
$B=26.7$, $V=26.8$, $R=26.0$, $I=25.7$, $z=24.3$, $J_s=23.9$, $K_s=22.5$.

To obtain reliable colours we must take into account that the imaging in the different wavebands was obtained under different seeing conditions. To derive and apply these corrections we did the following: first, we computed magnitudes in apertures of $2\arcsec$ in each of the available (``original'') waveband images. We then degraded the point spread function (PSF) of the original images to match the worst seeing condition and recomputed the $2 \arcsec$ aperture magnitudes in the ``degraded'' images. The corrections to be applied to the ``original'' magnitudes was then derived by comparing the magnitudes derived for bright stellar objects in the ``original'' and the ``degraded'' images.  

We used the Bayesian photometric redshift ({\sc bpz}) code of \cite{benitez2000} to derive photometric redshifts. The advantage of
the Bayesian approach is the use of {\emph{a priori}} probabilities by which it is
possible to include relevant knowledge, such as the expected shape of
redshift distributions and the galaxy type fractions, which can be
readily obtained from existing surveys. For our study we used the
same set of templates as those described in \cite{benitez2000}: four
\cite{coleman1980} templates (E/S0, Sbc, Scd and Irr), and
the spectra of two starburst galaxies in \cite{kinney1996}. We
derived two interpolated SEDs between each pair of these spectral
types. We used the priors derived by \cite{benitez2000} from the Hubble deep field north (HDFN) and Canada France redshift survey (CFRS) catalogues.

To quantify the reliability of the photometric redshift
estimation we used a sample of secure spectroscopic redshifts from our
campaign in this area \citep{specpaper}. In the region with 
multi-wavelength coverage described above, there are 120  such
spectroscopically identified sources.

\begin{figure}
\setcounter{figure}{1}
\resizebox{\hsize}{!}{{\includegraphics{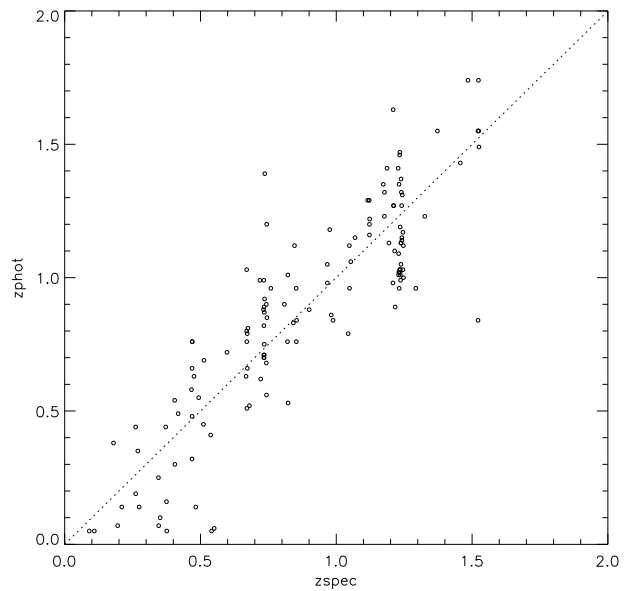}}}
\caption{Photometric redshift ($z_{phot}$) versus spectroscopic redshift ($z_{spec}$) for the full spectroscopic sample. Immediately recognizable is the cluster of galaxies at $z=1.237$. }
\label{zphzsp1}
\end{figure}

In  Fig.~\ref{zphzsp1} we plot $z_{phot}$ versus $z_{spec}$ for the spectroscopic sample. The photometric redshifts in general reproduce the spectroscopic redshift well but are systematically smaller. This is illustrated in Fig.~\ref{zphzsp2} where we plot the distribution of deviations of the photometric redshifts from the spectroscopic redshifts.  
The (sigma clipped) mean of the distribution is $\left<z_{phot}-z_{spec} \right >= -0.04$ and the standard deviation is  $\sigma  =0.18$.
\begin{figure}
\setcounter{figure}{2}
\resizebox{\hsize}{!}{{\includegraphics{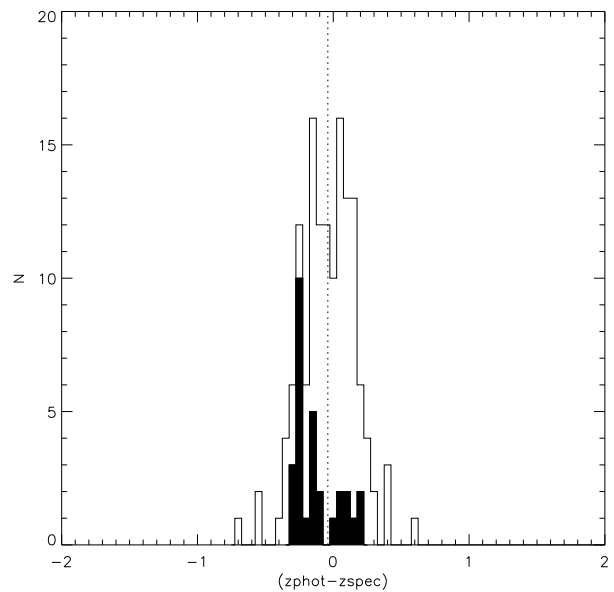}}}
\caption{Distribution of deviations of the photometric redshifts from the spectroscopic redshifts for the spectroscopic cluster member galaxies (filled histogram) and for the  full spectroscopic sample (open histogram).  }
\label{zphzsp2}
\end{figure}
If we restrict the analysis  to the spectroscopic cluster members  (the filled histogram in Fig.~\ref{zphzsp2}) the mean deviation is larger $\left<z_{phot}-z_{spec} \right >= -0.13$ and the standard deviation is $\sigma=  0.16$. 
In Fig.\ref{zphzsp3} we plot $\left<z_{phot}-z_{spec} \right>$ as a function of redshift to investigate whether this is a consequence of a redshift dependent systematic photometric redshift error, which could be introduced by small errors in the photometric zero-points of one or more of the filters. No significant dependency on redshift is observed. In the following we add 0.04 to all the derived redshift to empirically correct for the systematic deviation.

\begin{figure}
\setcounter{figure}{3}
\resizebox{\hsize}{!}{{\includegraphics{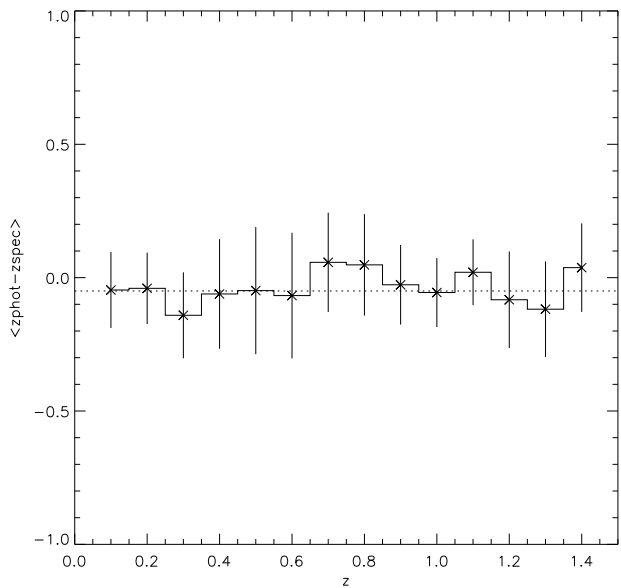}}}
\caption{Mean deviation of the derived photometric redshift from the spectroscopic redshift as a function of spectroscopic redshift, calculated in $\delta = 0.1$ bins. No significant dependency on redshift is observed. The dotted line marks the mean offset $-0.04$ of the full sample.}
\label{zphzsp3}
\end{figure}

To compare the photometric redshifts errors with those achieved in the literature we define for each object the reduced error in the photometric redshift estimation as $\delta _z = (z_{phot}-z_{spec})/(1+z_{spec})$. For the full sample we achieve a mean offset $\left < \delta _z \right >$=0.02 and an standard deviation of $\sigma (\delta_z)$=0.11. If we restrict our analysis to the cluster
members  we have $\left< \delta_z \right>$=0.05 and $\sigma (\delta _z)$=0.09.
The accuracy of our photometric redshifts are comparable to the accuracy achieved by \cite{barger2003} in a similar Bayesian photometric redshift study of galaxies in the Chandra deep field north. 
For the HDFN, \cite{benitez2000}  achieve a higher accuracy: $\sigma(\delta_z) = 0.06$, probably due to higher precession Hubble Space Telescope (HST) photometry.

\subsection{The photometric cluster member sample}
\label{sec.membersample}
In this section we describe how  the cluster galaxy population is
separated  from the foreground and background field galaxy population
using the photometric redshift. We take into account the uncertainty in the photometric redshift estimation by allowing galaxies with photometric redshifts within $\Delta z$ of the cluster redshift $z_{cl}$ to be classified as ``photometric members''.  The choice of $\Delta z$ is a trade off between cluster galaxy completeness and field galaxy pollution. The interval must be sufficiently broad to include as many ``real'' cluster galaxies as possible and sufficiently narrow to minimize pollution from field galaxies.
From Fig.~\ref{zphzsp1}-\ref{zphzsp3} it can be seen that most cluster member galaxies at $z=1.237$ are expected to have photometric redshifts in the range $\left|{z_{phot}-z_{cl}}\right | \le 0.3$.  Since we are aiming at building a sample of cluster galaxies which is complete to $K_s=22.5$ we prioritize not to exclude any cluster members and choose $\Delta z=0.3$. In the following, galaxies with $0.935<z_{ph}<1.535$ are thus classified as photometric members.

To estimate the ``completeness function'' of the photometric  member sample (the fraction of $z=1.237$ cluster galaxies recovered by the photometric redshift analysis in the $\Delta z$ interval, as a function of magnitude) we did the following:
for each $\delta m=0.5$ mag bin in the observed range of magnitudes $K_s=16-23$, we generated a catalog of 1000 galaxies at $z=1.237$ as they would appear in a dataset with the  same band-passes and limiting magnitudes as the data of \object{RDCS J1252.9-2927}. The catalog was generated using the {\emph{make\_catalog}} code in the {\sc {hyperz}} package \citep{hyperz} and were drawn randomly from 7 template spectral types (E to Im). 
We then applied {\sc {bpz}} to the catalogs. In Fig.~\ref{fig.completefun} we plot the fraction of  input ($z=1.237$)  galaxies with derived photometric redshifts within $0.935<z_{ph}<1.535$  as a function of $K_s$-band magnitude. Down to a magnitude of $K_s= 21.5$ about $90\%$ of the galaxies are recovered. At $K_s=22.5$ the recovery rate is $70\%$. 
The reason why {\sc bpz} appears to be performing slightly worse at the brightest magnitudes, compared to at fainter magnitudes, is that the simulated catalogs contain sizable fractions of bright late-type galaxies which are very rare at $z=1.237$, and therefore are assigned small probabilities by {\sc bpz}, resulting in a some cases, in the most likely spectrum being a lower redshift spectrum of earlier type.
\begin{figure}
\setcounter{figure}{4}
\resizebox{\hsize}{!}{{\includegraphics{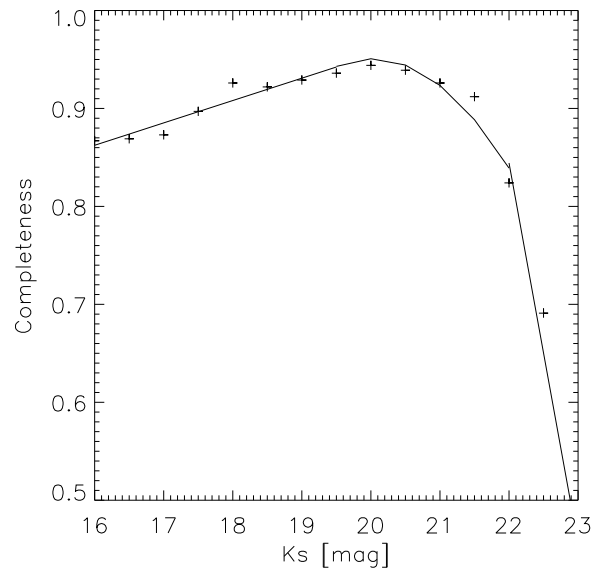}}}
\caption{
Completeness function of the photometric member sample, calculated as the fraction of simulated $z=1.237$ cluster galaxies recovered by the photometric redshift analysis as a function of magnitude.            
}
\label{fig.completefun}
\end{figure}
The redshift evolution of the field galaxy luminosity function, and its cosmic variance is not known with sufficient accuracy to correctly incorporate the effects of pollution of field galaxies in the completeness function analysis.  
We return to the issue of field contamination in \S\ref{sec.spatialdistribution} and \S\ref{sec.lumfun}.

\subsection{Spatial distribution of photometric member galaxies}
\label{sec.spatialdistribution}
To investigate the spatial distribution of the photometric members galaxies we constructed an image with the pixel value at the centroid of the photometric cluster members equal to one and the remaining pixel values equal to zero. This image was then smoothed with a Gaussian kernel with FWHM=60 pixels ($\sim8\farcs8$).       
In Fig.~\ref{imcont} we overlay contours of the smoothed density distribution of the photometric member galaxies on the $K_s$-band image of the cluster. The contours are $2-10$ times the density of galaxies in the HDFN with $K_s<22.5$ and   $0.935<z_{ph}<1.535$  \citep{HDFN}.

\begin{figure}
\setcounter{figure}{5}
\resizebox{\hsize}{!}{{\includegraphics{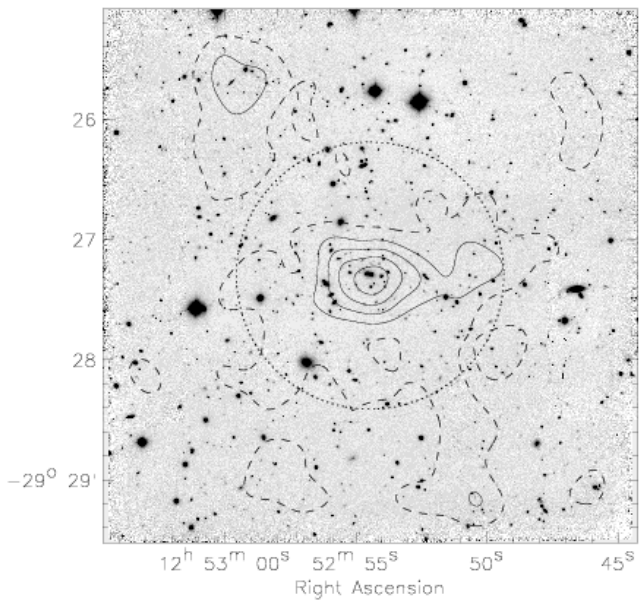}}}
\caption{Contours of the smoothed density distribution of photometric member galaxies overlaid on the $\sim 4\arcmin \times 4 \arcmin$ mosaic  $K_s$-band ISAAC image \citep{NIRpaper}. The contours are 2 times (dashed line) and 4-10 times (full line) the density of galaxies in the HDFN with $K_s<22.5$ and   $0.935<z_{ph}<1.535$ \citep{HDFN}. 
Inside the dotted circle with radius $65\arcsec$ the contamination from field galaxies in the photometric member sample is $25\%$. This circle marks the aperture inside which we derive the cluster galaxy luminosity function of \object{RDCS J1252.9$-$2927} in \S\ref{sec.lumfun}. }
\label{imcont}
\end{figure}

There is a large concentration of photometric  member galaxies in the central part of the field. In this region the photometric member sample must thus be dominated by cluster galaxies. The density of photometric members in the outer parts of the field is  comparable to the density in the HDFN  suggesting that field galaxy pollution could make a significant contribution to the photometric member sample in these regions. We discuss this further in \S\ref{sec.lumfun}

\begin{figure}
\setcounter{figure}{6}
\resizebox{\hsize}{!}{{\includegraphics{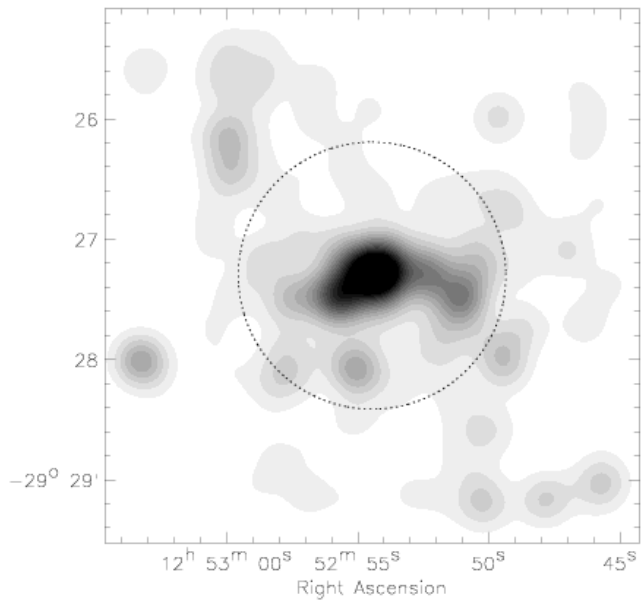}}}
\caption{Map of the smoothed $K_s$-band light of the photometric member galaxies. The map was generated by  creating a version of the K-band image in which pixels inside squares (of size 20 pix$\times$20 pix) centered on the photometric members were left at their original values and the remaining pixels were set equal to zero. This image was then smoothed with a Gaussian kernel with FWHM=60 pixels. The circle marks the region inside which the LF is derived.}
\label{Klight}
\end{figure}

In addition to the number-density distribution of the photometric member galaxies we also investigated the distribution of their $K_s$-band light. In Fig. \ref{Klight} we show a smoothed map of the $K_s$-band light of the photometric member galaxies. Since the $K_s$-band flux traces the stellar mass of the cluster galaxies, this map can be be thought of as a stellar mass map of the cluster, which can be compared to maps of the gas mass derived from X-ray observations, and maps of the dark matter derived from weak lensing observations. With this in mind it is  interesting to note that, in addition to a strong concentration of light in the center of the cluster, the light is extended in the East-West direction. This extension is also seen seen in both X-ray and weak-lensing mass maps \citep{weaklens}.   

\section{Luminosity function}
\label{sec.lumfun}

We now have a sample  of cluster galaxies which is complete to $K_s=22.5$, or rather a sample which we know how to correct for incompleteness, and we can derive the $K_s$-band luminosity function of the cluster galaxies without having to make uncertain statistical corrections to account for foreground and background field galaxy contamination. 
The choice of $\Delta z$ in \S\ref{sec.membersample} ensures that all spectroscopically confirmed cluster members are included in the photometric member sample.   
To take full advantage of the data we apply the maximum likelihood technique of \cite{schechter} directly to the luminosity distribution of the cluster galaxies rather than doing a ``least squares'' fit to a binned representation. 
In this way we include the information that in many magnitude intervals no galaxies are found, and do not make the assumption of the $\chi^2$ method that the underlying distribution is Gaussian. The ``incompleteness'' of the photometric redshift selection is taken into account through the completeness function shown in Fig.~\ref{fig.completefun}.  This in turn leads to more realistic error bars. For more details of the method we refer to the appendix of \cite{toft2016paper}. 

In \S\ref{sec.spatialdistribution} we argued that pollution from field galaxies in the photometric member sample could be significant in the outer parts of the $4\arcmin\times4\arcmin$ field.  The amount of pollution can be estimated from the subsample of photometric member galaxies with spectroscopic redshifts.  
These galaxies are likely to constitute a relatively fair sample of galaxies in the relevant redshift range, since the selection of the spectroscopic sample was designed not to introduce biases on the cluster galaxy populations, while minimizing the pollution of field galaxies. This was accomplished by targeting galaxies with $K_s<21$, $J-K_s<2.1$ and $R-K_s>3$. Such criteria do not penalize cluster galaxies, since at $z=1.237$ even the latest types are redder than $R-K_s=3$ and early types are bluer than $J-K_s=2.1$, however field contamination is significantly reduced. 

In Fig.\ref{fig.radiallumf} the bottom panel shows the contamination derived in apertures with increasing radius (centered on the cluster core).     
In the central parts of the field the contamination is modest. The pollution within  $45\arcsec$ is $\sim 10\%$, within $65\arcsec$ it is $\sim 25\%$, and within $120\arcsec$ it is $50\%$. 
To investigate the effects of the pollution on our analysis we derived the luminosity function of the photometric member sample in apertures of increasing size. The top and middle panel in Fig.\ref{fig.radiallumf} shows the variation of the luminosity function parameters with aperture radius.

There is a tendency for $K_s^*$ to be slightly brighter and $\alpha$ to be slightly smaller (less negative) in apertures encompassing only the central regions where the contamination is small,  compared to in larger apertures where the contamination is more pronounced, but the effect is barely significant since the error bars are larger in the smaller apertures due to the smaller number of galaxies. 
Part of the effect could be caused by intrinsic variation in the properties of the cluster galaxy LF with cluster centric distance, and part of it could be a consequence of field galaxy pollution. Since the effect is not statistically significant however, field galaxy pollution is not likely to significantly affect our results.      

Based on Fig.\ref{fig.radiallumf} we limit our cluster galaxy LF analysis to galaxies within $65\arcsec$ of the cluster center, in order to maximize the number of galaxies while minimizing the pollution.
The photometric member sample contains 100 galaxies within this distance, including 19  spectroscopically confirmed cluster members and 7 interlopers with spectroscopic redshifts different from $z_{cl}$. We remove the known interlopers from the sample and study the luminosity function of the remaining 93 galaxies in detail in the following.

\begin{figure}
\setcounter{figure}{7}
\resizebox{\hsize}{!}{{\includegraphics{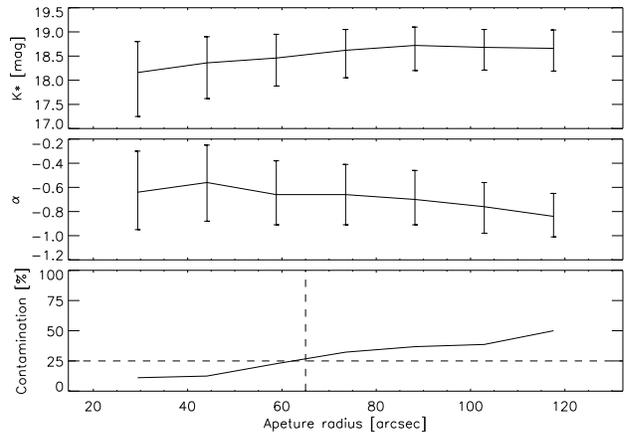}}}
\caption{The top and middle panel shows the $K_s^*$ and $\alpha$ parameters of the LF of the photometric member galaxies derived in apertures of increasing size, centered on the cluster core. Note that the two parameters are not independent, so the error bars in the two plots are correlated. The bottom panel shows the contamination in apertures of increasing size estimated from the subsample of photometric member galaxies with spectroscopic redshifts. The contamination is calculated as the number of interlopers, i.e. photometric member galaxies with spectroscopic redshifts different from the cluster redshift (defined as $1.22 \le z_{spec} \le 1.25$), divided by the total number of photometric member galaxies with spectroscopic redshifts. Inside the aperture with radius $65\arcsec$ (marked by the vertical dashed line), the contamination is 25\%.  }
\label{fig.radiallumf}
\end{figure}

The Schechter function provides a good fit to the data. In Fig.~\ref{contour} we plot $1-3\sigma$ likelihood contours of the two Schechter function parameters.  There is some degeneracy between the two parameters, but we are able to put firm constraints on both the characteristic magnitude $K_s^*=18.54^{+0.45}_{-0.55}$ and the faint end slope  $\alpha=-0.64^{+0.27}_{-0.25}$. Such accuracy is  unprecedented at these redshifts.    

Since our analysis is based on photometric redshifts which can be sensitive to   photometric errors, the results could potentially be affected by small changes in the photometry. 
To find out whether this is the case, we carried out a series of Monte Carlo simulations to investigate how sensitive the photometric redshift distribution is to the photometric errors. We generated 10 realizations of the full dataset, by randomly perturbing the photometry of the galaxies within their $1\sigma$ error bars. We then derived their photometric redshift distribution, defined a photometric cluster member sample, removed known interlopers and derived their LF in exactly the same way as for the original data set. 
The best fitting LF parameters of the perturbed datasets (represented by small symbols in Fig.\ref{contour}) all fall within the $1\sigma$ contour of the original dataset, indicating that the LF analysis is robust with respect to photometric perturbations, and that the effect of field galaxies scattering in and out of the $\Delta z$ interval as the photometry is perturbed is small, otherwise we would expect to see larger variations in the LF parameters. 

In Fig.~\ref{fig.lumfun} we plot the best fitting Schechter function and a binned representation of the data.  
For comparison, we plot the local $z$-band cluster galaxy LF, rather than the local $K$-band cluster galaxy LF, since the observed $K_s$-band corresponds roughly to restframe $z$-band at $z=1.237$.
Following the method of \cite{vandokkumfranx96} absolute $z$-band magnitudes $M_z$ can be related to the observed $K_s$-band magnitudes through: 
\begin{equation}
M_{z, AB}=K_{s, AB}-5*log(d_L/10)+2.5log(1+z)+\beta(H-K)_{AB},
\label{eq.Mz}
\end{equation} 
where $z=1.237$ is the redshift, $d_L$ is the luminosity distance in parsecs  and $\beta(H-K)_{AB}$ is a colour term to compensate for the fact that the redshifted $z$-band does not match the observed $K_s$-band  exactly. 
The basic assumption made to derive this expression is that the flux at the redshifted $z$-band can be related to the observed $H$ and $K_s$-band flux by $F_{\nu}({\nu_{z_{band}}}(z))=F_{\nu}(\nu_{H})^{\beta}F_{\nu}(\nu_{K})^{1-\beta}$.
We adopt the value $(H-K)_{AB}(z=1.25)=0.57$ predicted by the passive evolution models of \cite{kodama97} for an elliptical $L^*$ galaxy formed at $z_f=3$ \citep[the mean formation redshift of cluster ellipticals derived from their CM relation,][]{NIRpaper} and calculate  $\beta=0.35$ by assuming a simple power-law for the shape of the SED between $\nu_H$ and $\nu_K$.

Contamination from faint field galaxies is not likely to significantly affect the results derived for the cluster galaxy LF in the central part of the field, where the photometric member sample is dominated by bright galaxies, but we note that the main effect of such an contamination would be to overestimate the faint end slope, making the derived faint end slope a formal upper limit to the intrinsic faint end slope of the cluster galaxy LF.

\begin{figure}
\setcounter{figure}{8}
\resizebox{\hsize}{!}{{\includegraphics{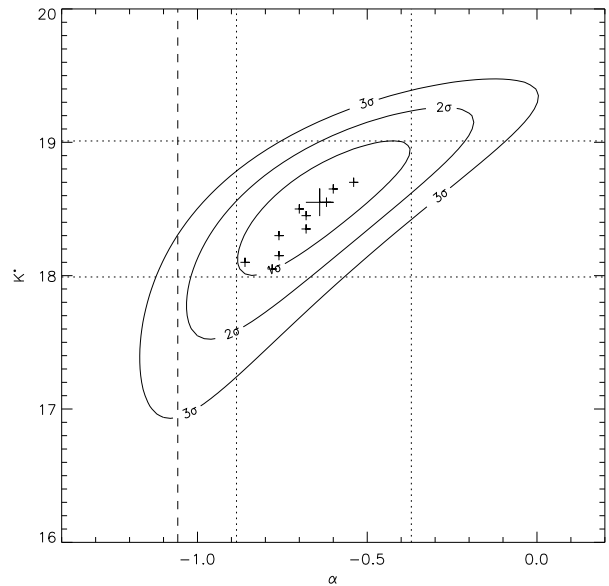}}}
\caption{Contour plot showing the constraints on the Schechter function parameters derived from the maximum likelihood analysis. The big cross marks the best fitting parameters, and the curves represent $1-3 \sigma$ confidence levels.
 The small crosses marks the best fitting parameters for the catalogs with perturbed photometry (see text). The dashed line mark the 1$\sigma$ upper limit for the faint end slope $\alpha=-1.14\pm0.08$ derived in local clusters at similar restframe wavelength \citep[in the $z$-band,][]{popesso2004}}
\label{contour}
\end{figure}

\begin{figure}
\setcounter{figure}{9}
\resizebox{\hsize}{!}{{\includegraphics{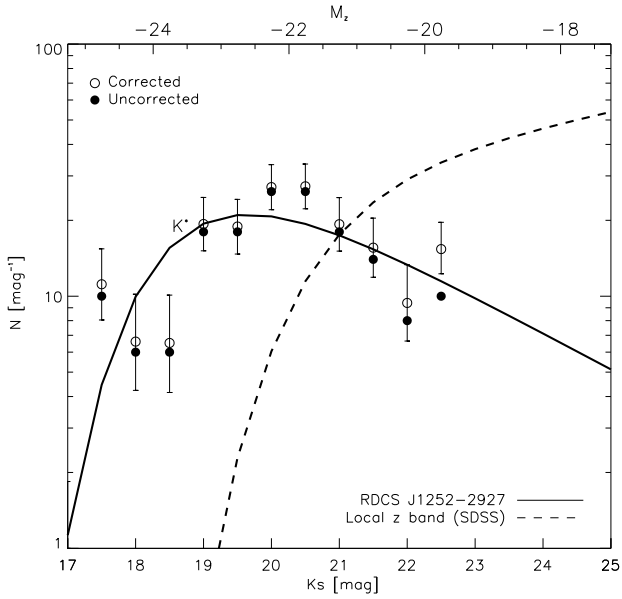}}}
\caption{
The full curve is the $K_s$-band LF of the photometric member galaxies within $65\arcsec$ of the cluster center, represented by the best fitting Schechter function, with parameters $K_s^*=18.54^{+0.45}_{-0.43}$ and $\alpha=-0.64^{+0.27}_{-0.25}$. 
The dashed curve is the local $z$-band cluster galaxy LF which has $M_z^*=-22.31\pm0.20$ and $\alpha=-1.14 \pm0.08$ \citep{popesso2004}. 
 The filled symbols are a binned representation of the raw counts, while the open symbols have been corrected for the incompleteness of the photometric redshift selection, using the completeness function in Fig.\ref{fig.completefun}. The conversion between observed $K_s$-band magnitude and restframe $z$-band magnitude is given in the text. }

\label{fig.lumfun}
\end{figure}

\section{LF evolution}
\label{sec.evolution}
The observed $Ks$-band  corresponds roughly to restframe $z$-band at $z=1.237$, so the evolution of the $z$-band LF can be constrained by comparing the observed $K_s$-band LF with the local $z$-band cluster LF.  
This is done in Fig.\ref{fig.lumfun} where we plot the derived $K_s$-band LF and the $z$-band LF of clusters in the local universe.
For the local LF we adopt the parameters $M_z^*=-22.31\pm0.20$ and $\alpha=-1.14\pm0.08$ derived by \cite{popesso2004} for the composite LF of X-ray clusters in the Sloan Digital Sky Survey (SDSS).    
From Fig.\ref{fig.lumfun} it can be seen that the characteristic magnitude of the restframe $z$-band LF at $z=1.237$ is brighter than at $z=0$. By transforming the observed $K_s^*$  to $M_z^*$ using Equation \ref{eq.Mz} an evolution of $\Delta M_z=-1.40^{+0.49}_{-0.58}$ mag is derived for the restframe $z$-band characteristic magnitude.
The observed $K_s^*$ can be converted to absolute (restframe) $K_s$-band magnitude by applying a k-correction:  
\begin{equation}
M_{K_s}=K_s-5*log(d_L/10) - k_{K_s}(z=1.237).
\end{equation}
If we adopt the   $k_{K_s}(z=1.25)=-0.68$ for the k-correction \citep{mannucci01} and compare to $M_K^*$ derived locally in the Coma cluster, we derive an evolution  in the restframe $K$-band characteristic magnitude of $\Delta M_K=-1.4\pm0.6$ mag (assuming that $M_K^*=M_{K_s}^*$), in agreement with the evolution derived in the restframe $z$-band.
The errors quoted for $\Delta M_z$ and $\Delta M_K$ are dominated by the uncertainty in the derived  $K_s^*$.
\begin{figure}
\setcounter{figure}{10}
\resizebox{\hsize}{!}{{\includegraphics{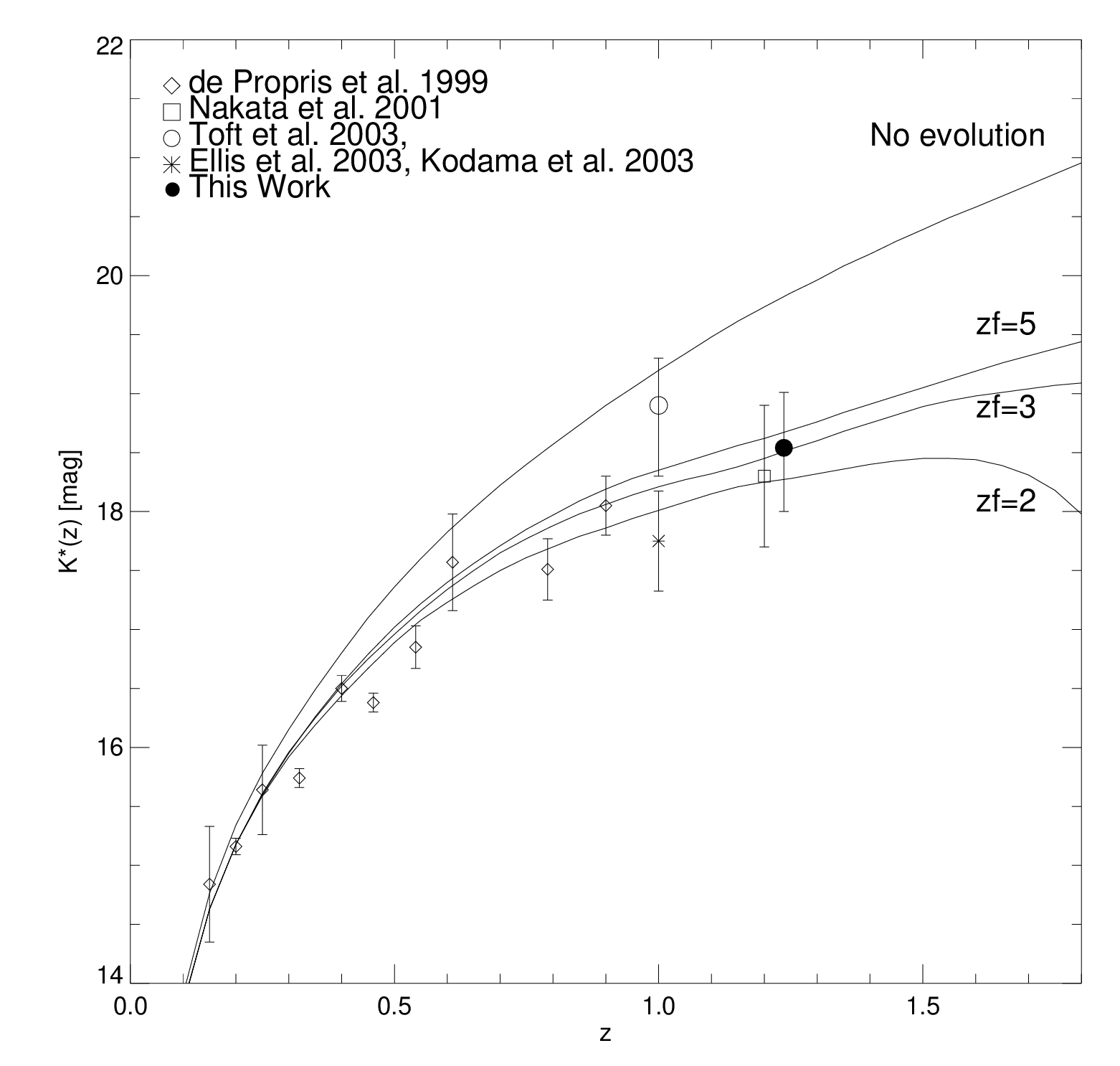}}}
\caption{Evolution of $K^*_s$ with redshift. The curves represent the evolution of an $L^*$ galaxy formed at $z_f=2$, 3 and 5 predicted by the passive evolution models of \cite{kodama97}, normalized to the Coma cluster which have $K^*_s=10.9$ \citep{depropris98coma}. The ``no evolution'' predictions are calculated using the k-corrections of \cite{mannucci01}. The data points from the literature  have been derived with a fixed $\alpha=-0.9$. \citep{depropris99, nakata01, ellis2004, kodama03} except for the $z=1$ value which was was derived with $\alpha$ as a free parameter \citep{toft2016paper}. The $z=1$ data point marked by a star, is the mean $K^*$ of 5 clusters in the redshift range $0.8<z<1.1$ \citep{kodama03,ellis2004}, with error bars added in quadrature.}
\label{fig.kstarevol}
\end{figure}
In Fig.\ref{fig.kstarevol} we compare the derived $K^*_s$ with values derived for clusters at lower redshift and the predictions of passive evolution models for the evolution of an $L^*$ galaxy formed  at $z_f=2$, 3 and 5 \citep{kodama97}. The observed evolution  of $K^*$ to $z=1.237$ is consistent with what is expected for a passively evolving population of galaxies formed at $z_f \ga 2$.

The derived faint end slope $\alpha=-0.64^{+0.27}_{-0.25}$ is similar to the value $\alpha=-0.60^{+0.39}_{-0.33}$ derived in the $K_s$-band in the $z=1$ cluster MG2016+112 \citep{toft2016paper},  but shallower (at the $2.2\sigma$ level, see Fig.\ref{contour}) than the value  $\alpha=-1.14\pm0.08$ derived at similar restframe wavelengths (in the $z$-band) in local clusters \citep{popesso2004}, indicating that clusters at $z\sim1$ contains relatively smaller fractions of low mass galaxies than clusters in the local universe. This is in disagreement with semi analytical hierarchical models which predict the faint end slope to steepen with redshift as the massive cluster galaxies break up into their progenitors.

\section{Summary and discussion}
\label{sec.discussion}
In this paper, we have taken advantage of an extensive NIR and optical dataset of the massive, X-ray luminous cluster of galaxies \object{RDCS J1252.9-2927} to derive the first secure constraints on the shape of the $K_s$-band LF at $z > 1$. 
The LF was found to be well represented by the Schechter function over the observed range of cluster galaxy magnitudes: $K_s=\left[17.0-22.5\right ]$. 

We tested our analysis for the influence of photometric errors and  pollution from field galaxies and found our results to be robust and relatively insensitive to the effects of field galaxy pollution in the central parts of the field where the constraints of the cluster galaxy LF is derived.

The  characteristic magnitude $K_s^*=18.54^{+0.45}_{-0.55}$ is $\Delta M_z=1.40^{+0.49}_{-0.58}$ magnitudes brighter than the characteristic magnitude measured for clusters in the local universe at similar restframe wavelengths (in the $z$-band). 

This is consistent with studies of the fundamental plane in a cluster at similar redshift where a luminosity evolution of $\Delta M_B=-1.50 \pm0.13$ was found  in the restframe $B$-band \citep{vandokkum03}. 

Apart from being shifted to systematically brighter magnitudes, the shape of the bright end of the LF at $z=1.237$ appears similar to in the local universe. 
Since the $K$-band LF is a good tracer of the stellar mass function of the cluster galaxies, this suggests that the massive elliptical that dominate the bright end of the LF were already in place at $z=1.237$. This is a challenge for hierarchical models which predict the bright end of the $K$-band LF to steepen and $K^*$ to become fainter at high redshift as the massive galaxies break up into their progenitors. At $z\sim 1$, current hierarchical models predict the characteristic mass (closely related to the characteristic $K$-band magnitude) to be a third of that in the local universe \citep{kodama03}, which is clearly not the case for \object{RDCS J1252.9-2927}.

The brightening of the characteristic magnitude, and lack of evolution in the shape of the bright end of the LF to redshift $z=1.237$ is  consistent with a simple formation scenario in which the massive elliptical galaxies that dominate the bright end of the $K$-band LF are passively evolving systems assembled at high redshift $z_f \simeq 3$.

This formation scenario is also in agreement with the observed properties of the CM-relation of elliptical galaxies in \object{RDCS J1252.9-2927}, which is identical to the CM relation found in local clusters in terms of slope and scatter, but bluer on average, consistent with old populations of stars formed at $2.7 < z_f < 3.6$ \citep{NIRpaper,blakeslee03}. From the evolution of the CM relationship alone it is not possible to distinguish between formation scenarios where the old stars are formed in monolithic collapse of the elliptical galaxies at high redshift, and scenarios where they are formed in the disks of less massive late-type galaxies which later merge to form the ellipticals, as long as the merging does not trigger significant star formation. 
From the lack of evolution in the shape of the bright end of the $K$-band LF we can however deduce that if the massive ellipticals in clusters formed through merging, it took place at higher redshifts ($z \gg 1$) than is predicted by current semi analytical models. 

The results derived here for the evolution of the bright end of the cluster galaxy $K_s$-band LF  are similar to the results derived from the K20 survey for the evolution of the field galaxy $K_s$-band LF \citep{pozzetti03}.  
The magnitude of the luminosity evolution ($\Delta M_K=-1.4\pm0.6$ mag to $z=1.237$) cannot be directly compared to the value derived in the field \citep[$\Delta M_K =-0.54\pm0.12$ mag to $z=1$,][]{pozzetti03} since the latter is a mean value derived in a broad redshift interval, centered on a lower redshift, but we note that they are broadly consistent.    

It is interesting to note that a population of  massive, evolved and highly clustered galaxies in the redshift range $2 \la z \la 4$ have recently been discovered in deep NIR observations \citep{franx2003,vandokkum2003b,daddi2003}. These are good candidates for progenitors of the old, massive, passively evolving elliptical galaxies observed in $z\sim 1$ clusters.    

The depth of our data allow us to trace the LF down to 4 magnitudes below $K^*_s$, which is unprecedented at these redshifts, and put firm constraints on the faint end slope of the LF.  
We derive a slope $\alpha=-0.64^{+0.27}_{-0.25}$ which is similar to the slope measured in the $K_s$-band in the MG2016+112 cluster at $z=1$ but shallower than the value measured at similar restframe wavelengths in clusters in the local universe. 
The observed evolution in the faint end slope can be interpreted as an evolution in the stellar mass spectrum of the low mass galaxies that dominate the faint end of the $K$-band LF, indicating that high redshift clusters contains relatively smaller fractions of low mass galaxies than clusters in the local universe. 
This results is in disagreement with semi analytical hierarchical assembly models, which predict the relative fraction of low mass galaxies to increase, and the faint end slope of the NIR LF to steepen with redshift as massive cluster galaxies break up into lower mass progenitors.

The results presented here are consistent with a scenario in which clusters are composed of a population of massive cluster galaxies which were formed at high redshift ($z\gg1$) and subsequently evolved passively with little additional star formation and interaction, and a population of lower mass galaxies which are continuously accreted from the field, primarily at lower redshift ($z<1$).

\begin{acknowledgements}
We thank T. Kodama for providing us with his elliptical galaxy evolution models and J. Hjorth for suggestions and discussions which helped improve the analysis and the presentation. We are grateful to the anonymous referee for very helpful comments.  
This work was supported by the Danish Ground-Based Astronomical Instrument Center (IJAF). 
Support for SAS came from NASA/LTSA grant NAG5-8430, who is also supported by the Institute of Geophysics and Planetary Physics (operated under the auspices of the US Department of Energy by the University of California Lawrence Livermore National Laboratory under contract W-7405-Eng-48).
\end{acknowledgements}

\end{document}